\documentstyle[12pt,epsfig]{article}
\voffset     = - 0.10 in
\begin{document}
\thispagestyle{empty}

\vspace*{1cm}

\begin{center}
{\Large\bf
On the Parameterization \\
of the Longitudinal Hadronic Shower \\[2mm] 
Profiles in Combined Calorimetry
}
\end{center}

\bigskip
\bigskip

\begin{center}
{\large\bf 
Y.A.~Kulchitsky$^{a, b,}$\footnote{
Correspondence address: 
Laboratory of Nuclear Problems,
Joint Institute for Nuclear Research, 
141980 Dubna, 
Moscow region, 
Russia.
Tel.: +7 09621 63782, 
fax:  +7 09621 66666; 
e-mail: Iouri.Koultchitski@cern.ch}, 
V.B.~Vinogradov$^{b}$} 

\bigskip

{\sl $^{a}$ Institute of Physics, National Academy of Sciences, Minsk, Belarus}
\\
\smallskip
{\sl $^{b}$ Joint Institute for Nuclear Research, Dubna, Russia}
\end{center}

\vspace*{1.0cm}

\begin{center}
{\it Submitted to Nuclear Instruments and Methods in Physics Research}\ 
{\bf A} \\[3.0mm]
{\bf Letter to the Editor}
\end{center}

\vspace*{\fill}

\begin{abstract}
The extension of the longitudinal hadronic shower profile parameterization 
which takes into account non-compensations of calorimeters and the algorithm 
of the longitudinal hadronic shower profile curve making for a combined 
calorimeter are suggested. 
The proposed algorithms can be used for data analysis from modern combined 
calorimeters like in the ATLAS detector at the LHC.
\end{abstract}
\newpage

One of the important questions of hadron calorimetry is the question
of the longitudinal development of hadronic showers.
This question is especially important for a combined calorimeter.

There is the well-known parameterization of the longitudinal hadronic shower 
development from the shower origin, suggested in \cite{bock81}. 
In \cite{kulchitsky98} this parameterization has been transformed to the 
parameterization from the calorimeter face
\begin{eqnarray}
        \frac{dE (x)}{dx} & = &
                N\
                \Biggl\{
                \frac{w X_0}{a}
                \biggl( \frac{x}{X_0} \biggr)^a
 e^{- b \frac{x}{X_0}}
                {}_1F_1 \biggl(1,a+1,
                \biggl(b - \frac{X_0}{\lambda_I} \biggr) \frac{x}{X_0}
                \biggr)
                \nonumber \\
                & & + \
                \frac{(1 - w) \lambda_I}{a}
                \biggl( \frac{x}{\lambda_I} \biggr)^a
                e^{- d \frac{x}{\lambda_I}}
                {}_1F_1 \biggl(1,a+1,
                \bigl( d -1 \bigr) \frac{x}{\lambda_I} \biggr)
                \Biggr\} \ ,
\label{elong03}
\end{eqnarray}
here ${}_1F_1(\alpha,\beta,z)$ is the confluent hypergeometric function, 
$X_0$ is the radiation length, $\lambda_I$ is the interaction length,
$N$ is the normalization factor; $a$, $b$, $d$ and $w$ are parameters:
$a = 0.6165 + 0.3193\ lnE$, $b = 0.2198$, 
$d = 0.9099 - 0.0237\ lnE$, $w = 0.4634$.
Note that the formula (\ref{elong03}) is given for a calorimeter 
characterizing by the certain $X_0$ and $\lambda_I$ values.
At the same time, the values of $X_0$, $\lambda_I$ and the $e/h$ ratios
are different for electromagnetic and hadronic compartments of a combined 
calorimeter. 
So, it is impossible straightforward use of the formula (\ref{elong03}) for 
the description of a hadronic shower longitudinal profiles in combined 
calorimetry. 

We have suggested the following algorithm of combination of the 
electromagnetic calorimeter ($em$) and hadronic calorimeter ($had$) curves 
of the differential longitudinal hadronic shower energy deposition $dE/dx$. 
At first, a hadronic shower develops in the electromagnetic calorimeter to 
the boundary value $x_{em}$ which corresponds to certain integrated measured 
energy $E_{em}(x_{em})$.
Then, using the corresponding integrated hadronic curve, 
$E(x)=\int_0^x(dE/dx) dx$, the point $x_{had}$ is found from equation 
$E_{had}(x_{had}) = E_{em}(x_{em}) + E_{dm}$.
Here $E_{dm}$ is the energy loss in the dead material placed between the 
active part of the electromagnetic and the hadronic calorimeters.
From this point a shower continues to develop in the hadronic calorimeter.
In principle, instead of the measured value of $E_{em}$ one can use the 
calculated value of $E_{em} = \int_0^{x_{em}} (dE/dx) dx$ obtained from the 
integrated electromagnetic curve. 
In this way, the combined curves have been obtained. 

These longitudinal hadronic shower develompent curves have been compared with 
the experimental data obtained by the combined calorimeter consisting of the 
lead-liquid argon electromagnetic part and the tile iron-scintillator hadronic
part \cite{comb96}.
This calorimeter has been exposed by the pion beams with energies of 10 -- 300
GeV. 

To reconstruct the hadron energy in longitudinal segments the new method of 
the energy reconstruction has been used 
\cite{kulchitsky99}.
In this non-parametrical method the energy of hadrons in a combined 
calorimeter is determined by the following formula:
\begin{equation}
	E  = 
		1 / e_{em}  \cdot (e/\pi)_{em}  \cdot R_{em}  + 
		1 / e_{had} \cdot (e/\pi)_{had} \cdot R_{had} + 
		E_{dm} \ ,
\label{ev7}
\end{equation}
here $R_{em}$ ($R_{had}$) is the electromagnetic (hadronic) calorimeter 
response, $e_{em}$ ($e_{had}$) is the electron calibration constants for 
the electromagnetic (hadronic) calorimeter.
The $(e/\pi)_{em}$ ($(e/\pi)_{had}$) ratio is
\begin{equation}
	\Bigl( \frac{e}{\pi} \Bigr)_{cal} 
		= \frac{( e/h)_{cal}}{1+((e/h)_{cal}-1)f_{\pi^0,\ cal}} \ ,
\label{ev5}
\end{equation}
where $cal = em,\  had$. 
For the electromagnetic and hadronic calorimeters the values of 
$(e/h)_{em} = 1.7$ and $(e/h)_{had} = 1.3$ are used.
The fraction of the shower energy going into the electromagnetic channel for 
electromagnetic compartment is $f_{\pi^0,\ em} =  0.11 \cdot ln{(E_{beam})}$.
The electromagnetic fraction in the hadronic calorimeter is equal to the one 
for shower with energy $E_{had}$: $f_{\pi^0,\ had} = 0.11\cdot ln{(E_{had})}$, 
where $E_{had} = 1 / e_{had} \cdot (e/\pi)_{had} \cdot R_{had}$.
This method uses only the known $e/h$ ratios and the electron calibration 
constants, does not require the previous determination of any parameters 
by a minimization technique, does not distort a longitudinal shower profile
and demonstrates the correctness of the reconstruction of the mean values of 
energies within $\pm 1\%$. 
Using this energy reconstruction method, the energy depositions $E_i$ have 
been obtained in each longitudinal sampling with the thickness of $\Delta x_i$
in units $\lambda_{\pi}$ 
\cite{comb96,kulchitsky99-1}.

\begin{figure*}[tbph]   
  \begin{center}
        \begin{tabular}{c}
\mbox{\epsfig{figure=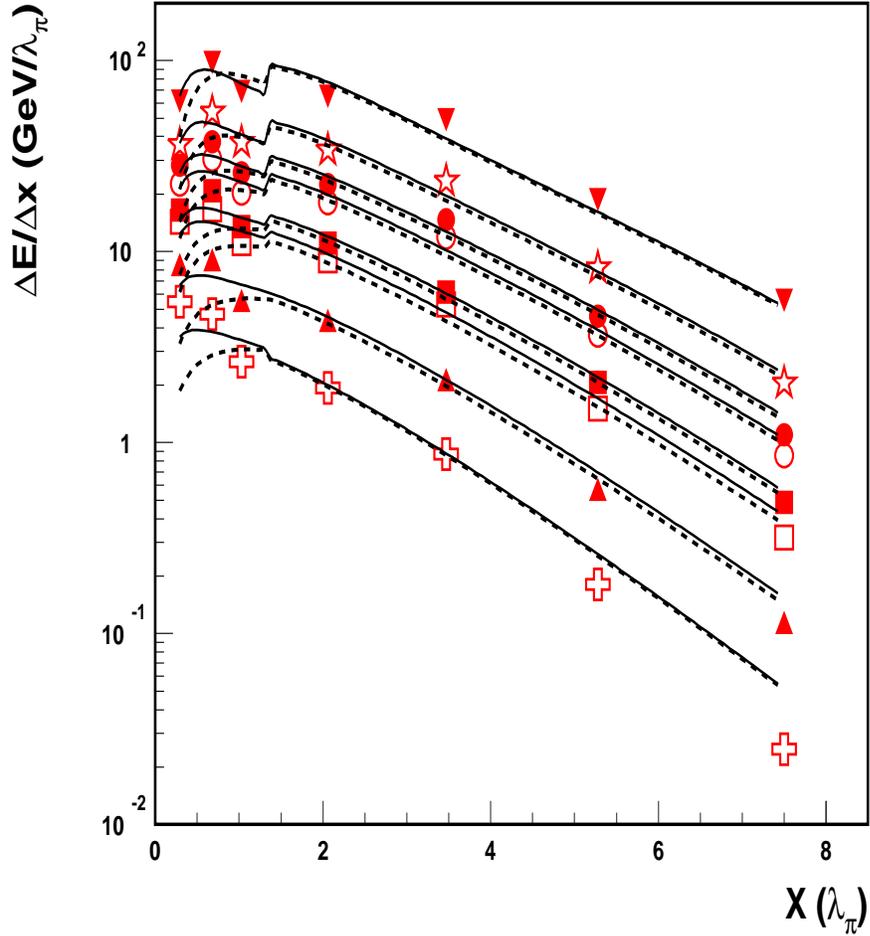,width=0.95\textwidth,height=0.85\textheight}}
        \\
        \end{tabular}
\vspace*{-20 mm}
\caption{
	The experimental differential longitudinal energy depositions at
	10 GeV (crosses), 20 GeV (black top triangles), 40 GeV (squares),
        50 GeV (black squares), 80 GeV (circles), 100 GeV (black circles),
        150 GeV (stars), 300 GeV (black bottom triangles)
        energies as a function of the longitudinal coordinate $x$ in units
        $\lambda_{\pi}$ for the combined calorimeter and the results 
        of the description by the Bock et al.\ (dashed lines) and modified 
	(solid lines) parameterizations.
	}
\label{fv6-2}
     \end{center}
\end{figure*}

Fig.\ \ref{fv6-2} shows the differential energy depositions
$(\Delta E/ \Delta x)_i = E_i / \Delta x_i$ as a function of the longitudinal 
coordinate $x$ in units $\lambda_{\pi}$ for the 10 -- 300 GeV and comparison 
with the combined curves for the longitudinal hadronic shower profiles
(the dashed lines).
It can be seen, there is a significant disagreement in the region of the 
electromagnetic calorimeter and especially at low energies.

We attempted to improve the description and to include such essential feature 
of a calorimeter as the $e/h$ ratio.
Several modifications and adjustments of some parameters of this 
parameterization have been tried.
It turned  out that the changes of two parameters $b$ and $w$ in the formula 
(\ref{elong03}) in such a way that 
\begin{equation}
b = 0.22 \cdot (e/h)_{cal} / (e/h)_{cal}^{\prime}\ ,
\label{eq-b}
\end{equation}
\begin{equation}
w = 0.6  \cdot (e/\pi)_{cal} / (e/\pi)_{cal}^{\prime} 
\label{eq-w}
\end{equation}
made it possible to obtain the reasonable description of the experimental data.
Here the values of the $(e/h)_{cal}^{\prime}$ ratios are 
$(e/h)^{\prime}_{em} \approx 1.1$ and $(e/h)^{\prime}_{had} \approx 1.3$
which correspond to the data used for the Bock parameterization \cite{bock81}.
The $(e/\pi)_{cal}^{\prime}$ are calculated using formula (\ref{ev5}). 

In Fig.\ \ref{fv6-2} the experimental differential longitudinal energy 
depositions and the results of the description by the extension of the 
parameterization (the solid lines) are compared.
There is a reasonable agreement (probability of description is more than
5\%) between the experimental data and the curves, taking into account 
uncertainties in the parameterization function.

So, we propose the extension of the longitudinal hadronic shower profile 
parameterization which takes into account non-compensations of calorimeters and
the algorithm of the longitudinal hadronic shower profile curve making 
for a combined calorimeter. 

We are thankful Peter Jenni, Marzio Nessi and Julian Budagov for fruitful 
discussions and support of this work.



\clearpage

\end{document}